\documentstyle[prl,aps,preprint]{revtex}
\draft
\begin{document}
%\wideabs{
\title{
Self Organized Criticality in Digging Myopic Ant Model.}
\author{Prashant M. Gade$^1$\cite{emprash} and M. P. Joy$^2$\cite{emjoy}}
\address{$^1$Jawaharlal Nehru Centre for Advanced Scientific Research, 
Jakkur, Bangalore-560064,  INDIA\\
$^2$Materials Research Centre, Indian Institute of Science,
Bangalore-560012, INDIA}
\maketitle
\begin{abstract}
We demonstrate the phenomenon of self organized criticality (SOC)
in a simple random walk model described by a random walk of a
myopic ant. The ant acts on the underlying lattice aiming at
uniform digging of the surface but is unaffected by the underlying lattice.
In 1-d, 2-d and 3-d we have explored this model and have obtained
power laws in the time intervals between 
consecutive events of `digging'. Being a simple
random walk, the power laws in space translate to power laws in
time. We also study the finite size scaling of asymptotic
scale invariant process 
as well as dynamic scaling in this system. This  model 
differs qualitatively from the cascade models
of SOC. 
\end{abstract}
%}

The concept of self organized criticality (SOC) was introduced by
Bak, Tang and Wiesenfeld in the context of avalanches in sandpile
model (BTW model)\cite{BTW}. Diffusively coupled spatially extended system 
which is driven adiabatically, i.e. the drive occurs only when system
has been fully relaxed, settles in the metastable 
state with very long correlations
and no characteristic length scale. This model is termed to be self
organized since  the critical state is  reached, though no particular
parameter seems to have been adjusted. There have been  further variants 
of the above model which have  similar rules, but are in different 
universality class \cite{Manna}. 
The above models are cellular automata models in which the discrete
variable value assigned to different points on a d-dimensional  lattice
are updated in discrete time \cite{Zhang}. 
The relevant perturbations
 in which SOC gets destroyed has been a topic
of interest to  many researchers \cite{Ram}.  
Developing a PDE model for SOC has also been an active area of interest
 \cite{Grin}.
There have been models with threshold dynamics in continuous variable
values like adaptive dynamics model on coupled map lattices or earthquake
models, though it is debatable whether the power laws arising in these
models can be termed as self organized \cite{SS,Carlson}. 

In all these models SOC is induced by a branching process. The disturbance
propagates from one length scale to the other 
by branching
in various directions and this hierarchical basis for
the dynamics leads to a power-law
behavior. This description of branching leading to power laws has
been given for as diverse processes as
intermittent turbulent process by Kolmogorov \cite{Kol}
or
income distributions in US by Schlesinger \cite{Sch}. However, scale invariant 
processes need not be produced by branching alone. The disturbance can choose
a random direction yielding scale invariant structure. 
Here we propose a simple random walk model for SOC.   
As a physical illustration,  we would like to note a recent
experimental 
observation by Vishwanathan {\it et al} \cite{Nature} about foraging
behavior of sea-birds. In this experiment, 
the authors studied the foraging behavior of wandering albatross.
Measurements of the distance travelled by the
bird at various times are carried out.
They found a power law behavior in distribution of flight time events. 
Interestingly, the observation is that though the distribution deviates
significantly from simple random walk,  it is still a 
power law implying a scale invariant manner in which the 
flights proceed.  Assuming that the flight directions change randomly after 
finding food, they  argued that the data they have suggests that
the distribution of food on the  ocean surface is also  
scale invariant. Although we do not attempt to model this experiment, it 
nicely illustrates the fact that
not only the branching processes but  the 
processes induced by simple random walk/flight also  can organize themselves
in a scale invariant fashion in time and space. 

We introduce a new model of self organized scale invariant
behavior in space and time which is induced by random walk. 
A model of Eulerian walkers(EW) has been
introduced recently \cite{Dhar}. Our model is 
simpler in the sense that unlike the above model, the walker is 
unaffected by the medium. As it  will be clear in the course of discussion,
not only it is in a different universality class, but is even qualitatively
different from the earlier models.

Let us first 
discuss our model in 1-dimension for simplicity. We consider
a lattice of length $L$. At each site $i$, $1\leq i \leq L$, we associate
an integer, $x_i$ which denotes the height of that 
point and $-\infty < x_i \leq 0$. To begin with,
we assign $x_i = 0$ for all $i$.
We put a random walker at a randomly
chosen site $j$, ($1\leq j \leq L$). Now the dynamics of the  lattice is 
defined in the following way. (a) At each time step, the  random walker moves
to its nearest neighbor which is chosen randomly. (b) Before moving to the next site,
the  random walker compares the height at that site 
with those of nearest
neighbors and   reduces  the height at that 
site by 1 unless any of the nearest neighbors has a higher height.
In  other words, if random walker is at site $k$, then
\[
x_k=x_k-1 \;\;{\rm unless}\;\; x_{k+1}> x_k \;{\rm or}\; x_{k-1}>x_k.
\]
We will note this event of reduction of height as `digging'. 
The condition above on digging is introduced 
since the aim of the random walker is to dig uniformly
and it does not want to dig the site which already has a height lower 
than any of its neighbors. Though the medium is affected by the walk,
the walker is unaffected by the medium, i.e. the next site to which
the random walker moves is chosen randomly and is independent of the
entire height profile.
At boundaries the comparison is only one sided.
If the random walker moves out of the lattice, we put it back in a 
randomly chosen site within the lattice. Since the random walker
can not see beyond nearest neighbors we call the model
as digging myopic ant (DMA) model. 
We can also describe the model in terms of the evolution rule for the
slopes on either side of the random walker. (By construction the 
slopes can take only three values 1,0, and -1.) Of the nine
possible combination, four of them transform as
$1,0\longrightarrow 0,1;\, 1,-1 \longrightarrow 0,0;\, 0,0 
\longrightarrow -1,1;\, 0,-1 \longrightarrow -1,0$ while 
the rest five  remain unchanged. (See Fig. 1.)
The rule at the left boundary is $0 \longrightarrow 1;\, -1\longrightarrow
 0$ and $1 \longrightarrow 1$ while at the right boundary $1 \rightarrow 0$,
$0\rightarrow -1$ and $-1\rightarrow -1$. 
Note that except at boundaries the sum of slopes remains
conserved. 

We start with a flat surface. This means that in the beginning all sites 
are potentially `active', i.e. can be dug. 
However, as the surface evolves, all kinds
of valleys appear in the interface and  only a few sites at the top of
the valley 
remain active. If one ignores the fact that the sites dug subsequently
are not independent of each other, rather are spatially nearby, i.e. 
the noise in our case is correlated, one can relate the
distribution of times required to reach active sites to
the spatial distribution of active sites. Now we look at the 
distribution of time intervals between which active sites were visited.

Here, the 
system is driven by random perturbations and the
time interval $t$ between two successive events of digging when the
medium is affected by the walker is a quantity of interest.
We compute the distribution  $P(t)$ where $P(t)$ is the 
normalized probability that the time between two successive
events of digging is $t$. We also
compute the 
probability distribution  $D(s)$,
 the number of distinct sites $s$ visited by
the random walker between two successive events. We find that
$P(t) \sim t^{-\gamma},\; \gamma\approx 1.6$. 
It is clear that the distribution $D(s)$
can not be independent of $P(t)$ since in a simple random walk number
of distinct sites $s$ visited in time $t$ goes as $t^{{\frac{1}{2}}}$.
This implies $D(s)\sim s^{-\gamma'},\; \gamma'=2\gamma -1$. 
Thus as one would expect, a
power law distribution in time translates in a power law distribution in
space. Fig. 2(a) and Fig. 2(b) show $P(t)$ and $D(s)$ for
 various lattice sizes in 1-d.
If one looks at the spatial profile of the lattice developed after a long
time, one can see valleys of all sizes. Thus a myopic random walker who 
started the walk aiming at a uniform digging of the surface, ends up
digging the surface in a scale invariant manner. 
Thus unlike BTW model,  this model shows
nontrivial nontransient scaling properties even in one dimension.
However, we note that
it has properties common with earlier SOC models. It is a
conservative model except at the boundaries 
in the sense that the sum of slopes at all the sites
does not change unless digging occurs at the boundary. As in
earlier model,  the boundary 
conditions are open. However, as seen above, evolution
rule described in terms of local slopes is anisotropic.
The relation with the distribution of active sites
is  not clear since noise is correlated.

Given the nature of the distributions, i.e. a simple power law followed
by an exponential tail, one can fit a finite size 
scaling form   
 $P(t,L)=L^{\mu}G(t/L^{\nu}), D(s,L)= L^{\mu'} F(s/ L^{\nu'})$,
($\mu=\gamma \nu$ and $\mu'=\gamma' \nu'$), to
the distributions \cite{Kada}.
In 1-d we can fit the scaling nicely with $\nu=2,\nu'=1$.
This is useful in higher dimensions in particular where
it is difficult to do a very large size simulations and scaling form 
gives the power law exponents with reasonable accuracy.
In the Fig. 2 depicting the distributions $P(t)$ and $D(s)$,  
we  also show the finite-size scaling in the inset.

The model can be easily extended to  higher dimensions. We have studied this
model in two and three dimensions. We plot inter-event time distribution 
 $P(t)$ in 2-d and 3-d in Fig. 3(a) and Fig. 4. 
As in 1-d, $P(t)\sim t^{-\gamma}$ 
with $\gamma\approx  1.2, \nu\approx 2$ in 2-d and $\gamma\approx 1.2, \nu
\approx 1.8$ in 
3-d. Since the number of distinct sites covered
$s$ goes as $t/ln(t)$ in 2-d and as $t$ in 3-d\cite{Nemi}, 
one can expect a power law 
distribution for $D(s)$ as well with $\gamma'=\gamma$
except that one expects a logarithmic
correction in 2-d (which was not possible to detect for the sizes
to which we could carry out the simulations). Fig. 3(b)
 shows the distribution  
$D(s)$ in 2-d. For 3-d, site distribution
was beyond our available computational resources. However, we expect
it to closely follow the $P(t)$. In Figures 3 and 4, the insets show 
the  finite size scaling in each of the cases as  in Fig. 2.
The geometrical picture in 2-d is identical to that in 1-d. One sees valleys
of all sizes present in the asymptotic height profile in 2-d.  
This is understandable. Like in sandpile model if one has a configuration
with a single big valley, the random walker can go to the boundary and
dig  making  sites in the interior active and thus
one expects many events. (In our model, one more configuration in which
not many sites will be active will be a long tilted interface. However,
by the same logic, it will not stay for long.)
Similarly, starting with a flat interface,
one expects many events since all sites are active. Thus the
surviving configuration, or the configuration which will be attained
most of times will be the one in which valleys of all sizes are present.

We have also seen how the profile changes in time. The 
simplest quantitative measure that demonstrates the geometrical changes 
in the profile is roughness. The roughness $\sigma(L,t)$ of the
interface of length $L$ at time $t$ (starting with a flat interface)
is given by $\sigma(t,L)=\sqrt{\frac{1}{L}
\sum_{i=1}^L(x_i(t)-\bar x(t))^2}$, where $\bar x(t)$ is the average
height of the interface at time $t$. Growth depends 
on nearest neighbors and thus the correlations develop in time
and span the entire length $L$. When the entire surface gets correlated
the width saturates. The roughness $\sigma(L,t)$ follows a scaling
relation, $\sigma(t,L)=L^\alpha f(t/L^{z})$ (See  {\it e.g.} \cite{stan}). 
The exponent $z=\alpha/\beta +1$. The exponent $\beta= 0.565$ 
signifies the growth in time in the begining ($\sigma(t,L) \sim t^\beta$),
$z$ gives  saturation 
time ($t_{sat}  \sim L^z$) and $\alpha=1.1$ signifies saturation width
($\sigma_{sat} \sim L^\alpha$). The scaling form with the above fit which
assumes a power-law growth followed by saturation is reasonably good 
(see Fig. 5).  For small 
times ($t<9,\; L>>t$) one can easily compute all the possible configurations
and their probabilities analytically. The values computed so are in close 
agreement with simulations and also yield the growth exponent
$\beta=0.565$.  Large value of $\alpha$ reflects the highly inhomogeneous
asymptotic interface. 

We have also studied a variant of the
model in which one tries to reduce the correlation between successive
events by putting the random walker in a random position after 
each digging. Thus the noise is not spatially correlated any longer.
Most of the qualitative features
of the model do not change. 
The dynamic scaling in this variant and further investigations 
in the current model as well as its variant are deferred
to a future publication.

In short, we have proposed a new model of self organized criticality
in which the governing mechanism is that of diffusion. This model is
hopefully easier to handle analytically since the exponents in
space are easily related to exponents in time and one does not
have a lot of unrelated and and ill-understood exponents.  
We also feel that such models could be of use in situations 
which yield scale invariant behavior but do not involve cascades,
but rather have diffusion as the only way in which information spreads
in the system.

Authors have enjoyed discussions with  N. Kumar
and  G. Ananthakrishna. PMG acknowledges correspondence 
with D. Dhar. MPJ acknowledges financial support
from IFCPAR Grant no.1108-1.

\begin{figure}
\caption{
Schematic diagram of the configurations that change by the action of
the random walker which is  at the centre. Slopes are also shown.
}
\label{figure1}
\end{figure}
\begin{figure}
\caption{
(a)The inter-event time distribution, $P(t,L)$ {\it vs} 
time $t$ in 1-d. 
(b)Probability distribution $D(s,L)$ that
$s$ distinct sites are visited between two events,  {\it vs}
  $s$
in 1-d. (In both figures  $A)\, L= 10,\;$ $ B)\, L= 25,\;$
$ C)\, L=50,\;$ $ D)\, L=100$ and
$E)\, L=1000$. Insets show finite size scaling.) 
}
\label{figure2}
\end{figure}

\begin{figure}
\caption{
(a)$P(t,L)$ {\it vs}  
time $t$ in 2-d.
(b)$D(s,L)$ {\it vs}  $s$
in 2-d. (In both figures $A)\, L=10, \;$ $ B)\, L=25,\;$ $
 C)\, L=50, \;$ $ D)\, L=100 $ 
and $ E)\, L=200$. 
Insets show finite size scaling.)
}
\label{figure3}
\end{figure}

\begin{figure}
\caption{
$P(t,L)$ {\it vs}  $t$ in 
3-d.
Inset shows finite size scaling.
}
\label{figure4}
\end{figure}

\begin{figure}
\caption{
 Roughness $\sigma(t,L)$ {\it vs}
time $t$ for various $L$ in 1-d. Inset shows
the dynamic scaling of  
the
interface. 
%The symbols correspond to different system sizes,
%$L=20 \; (\diamond), \;  L=40 \; (+), \;
% L=80 \; (\Box),
% \;  L=160 \; (\times)$ and $L=320 \; (\bigtriangledown)$. 
}
\label{figure5}
\end{figure}
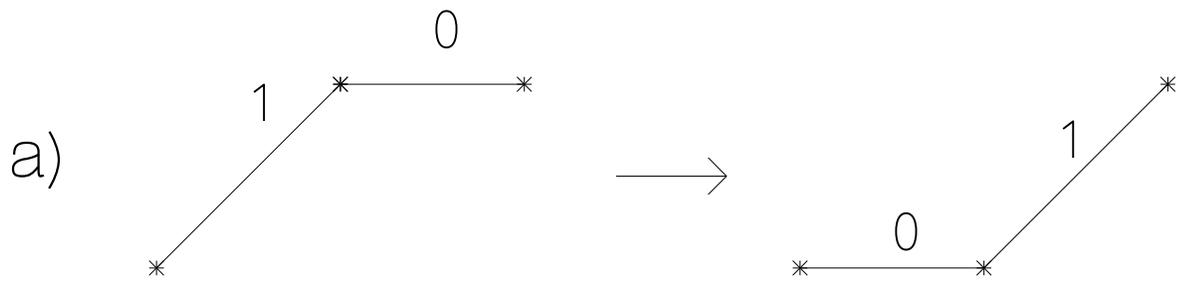
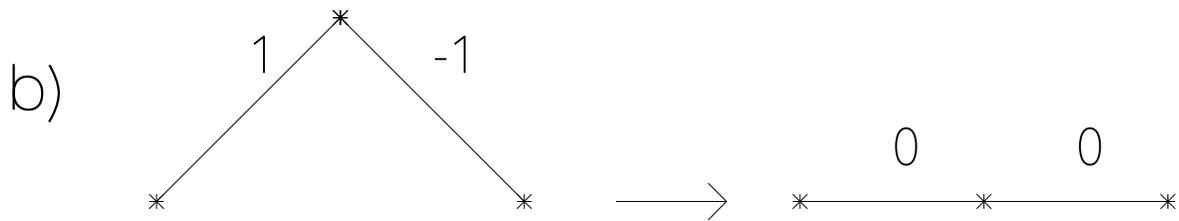
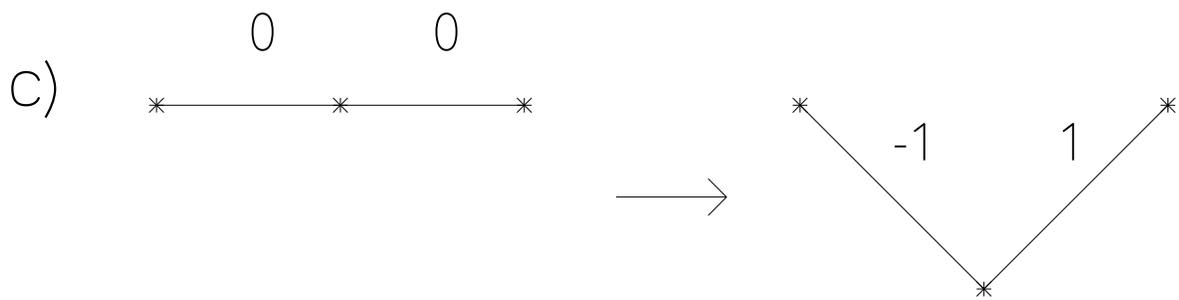
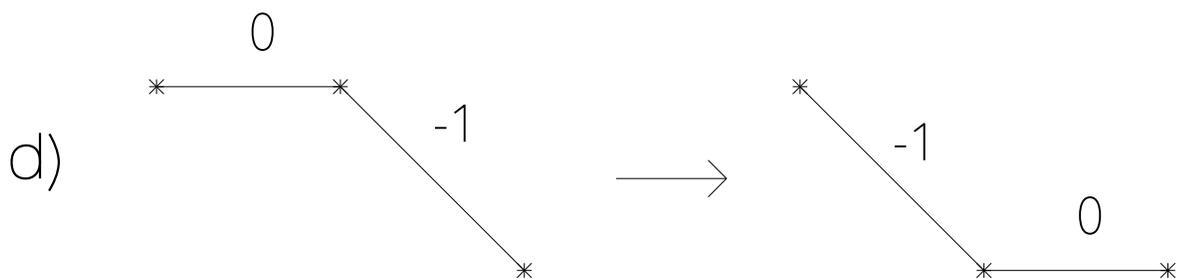
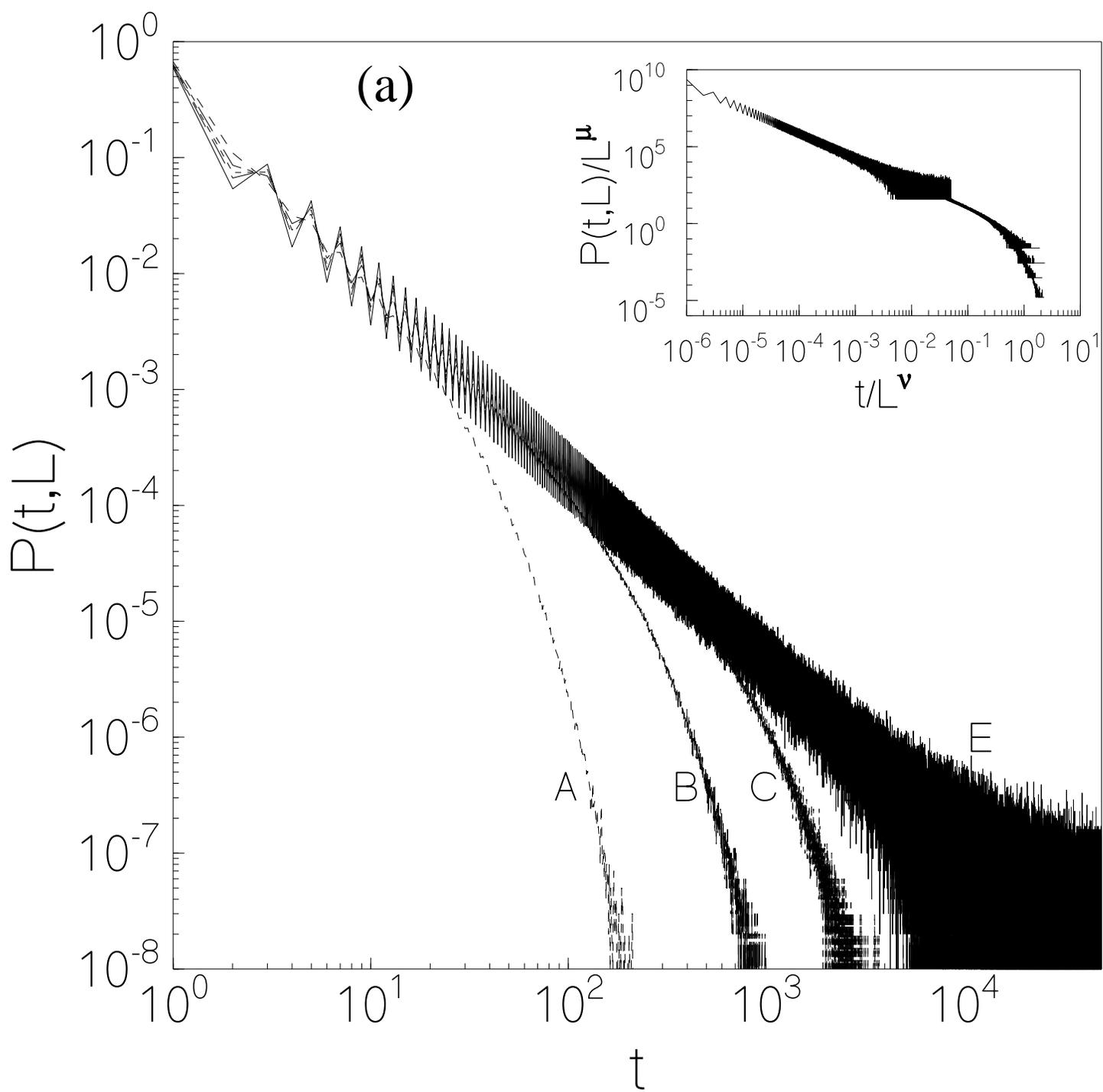
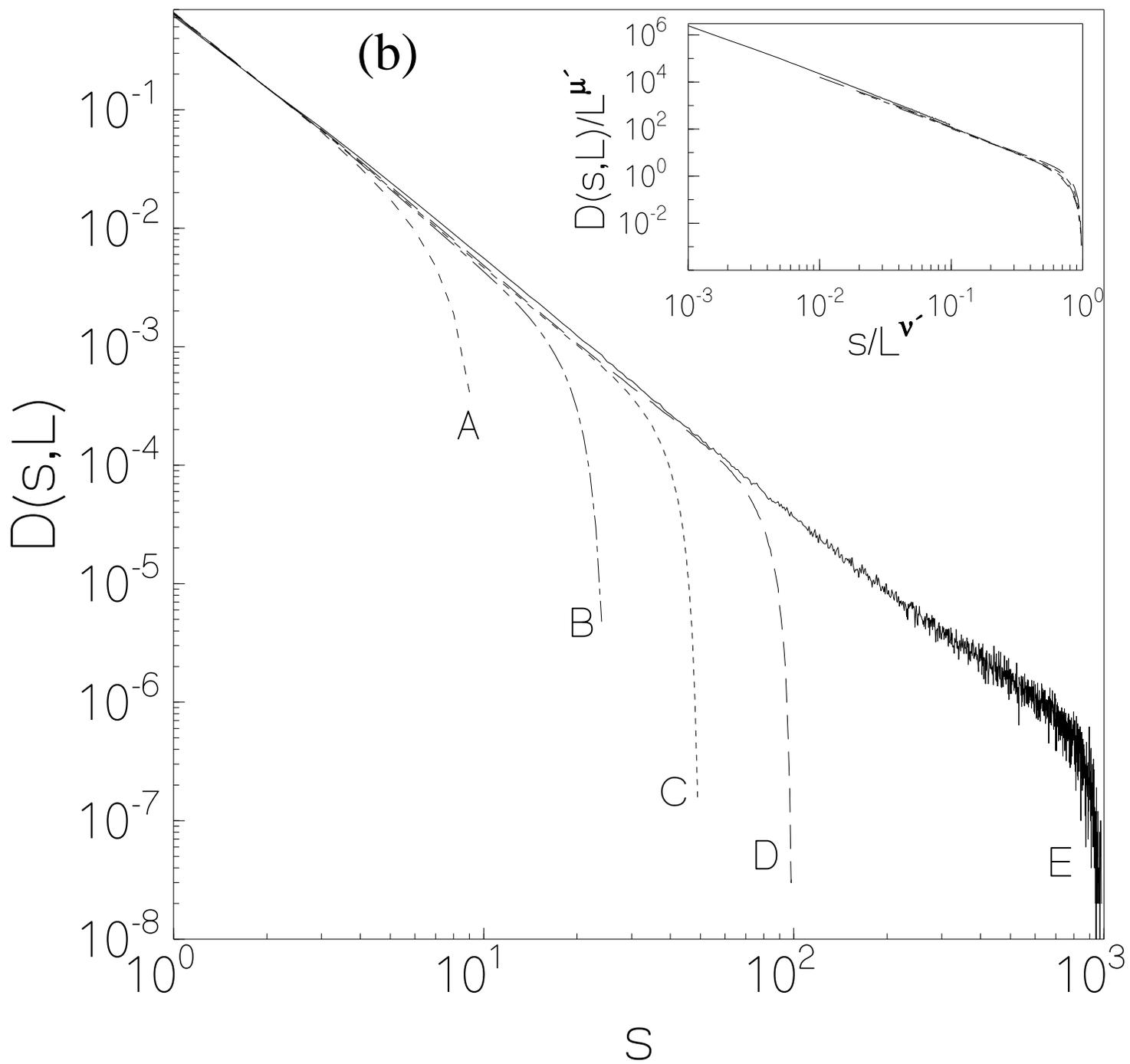
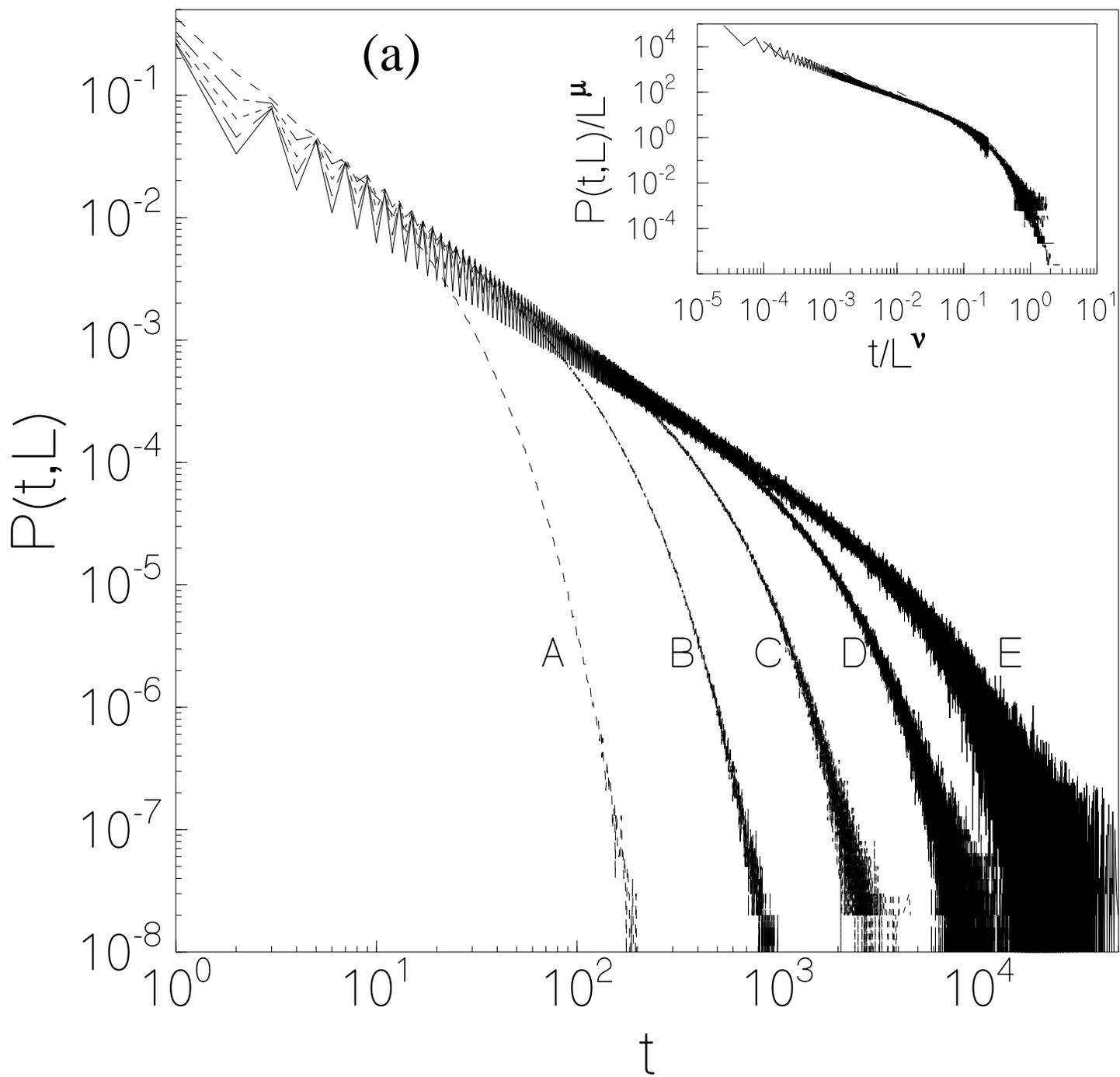
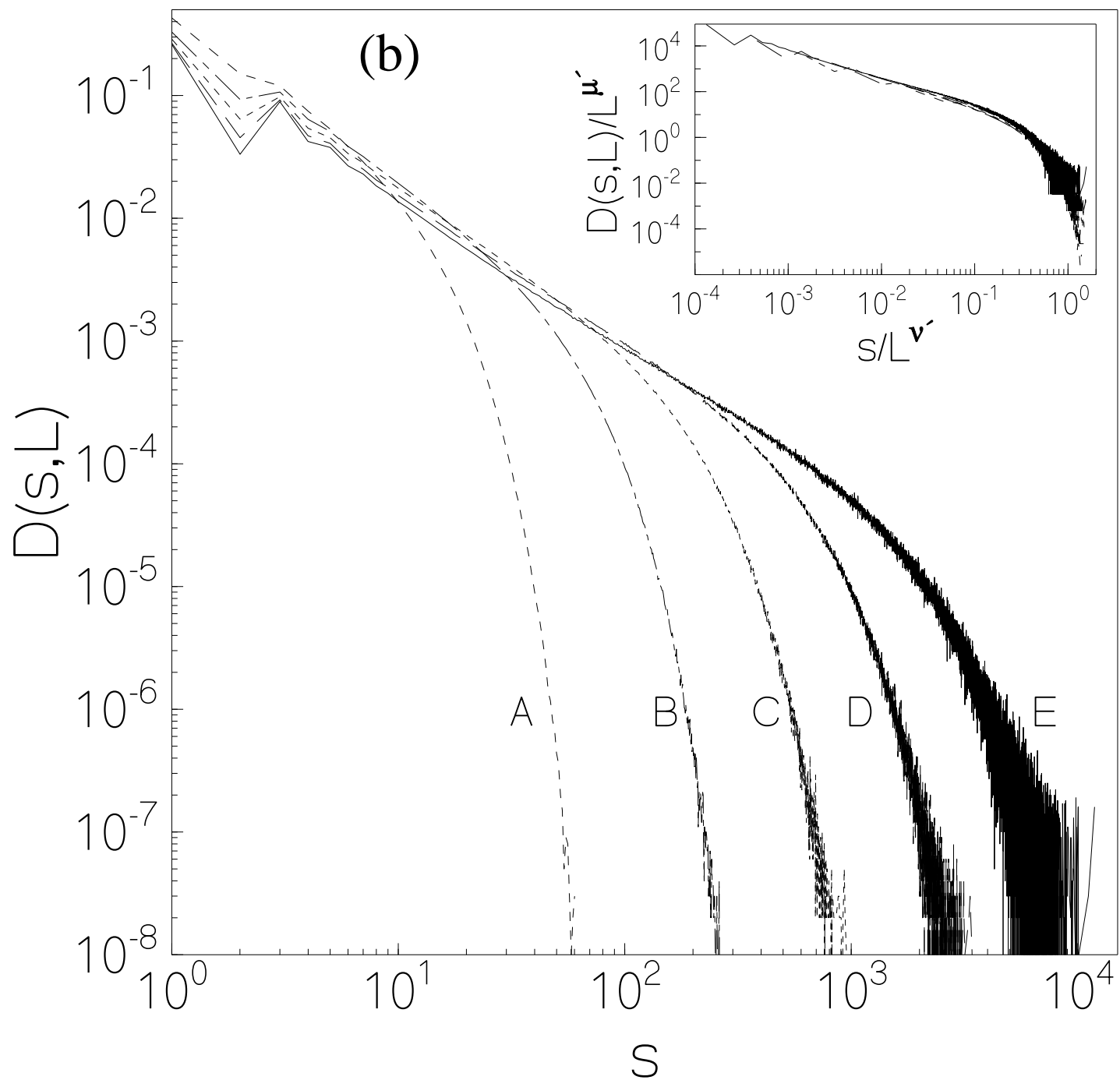
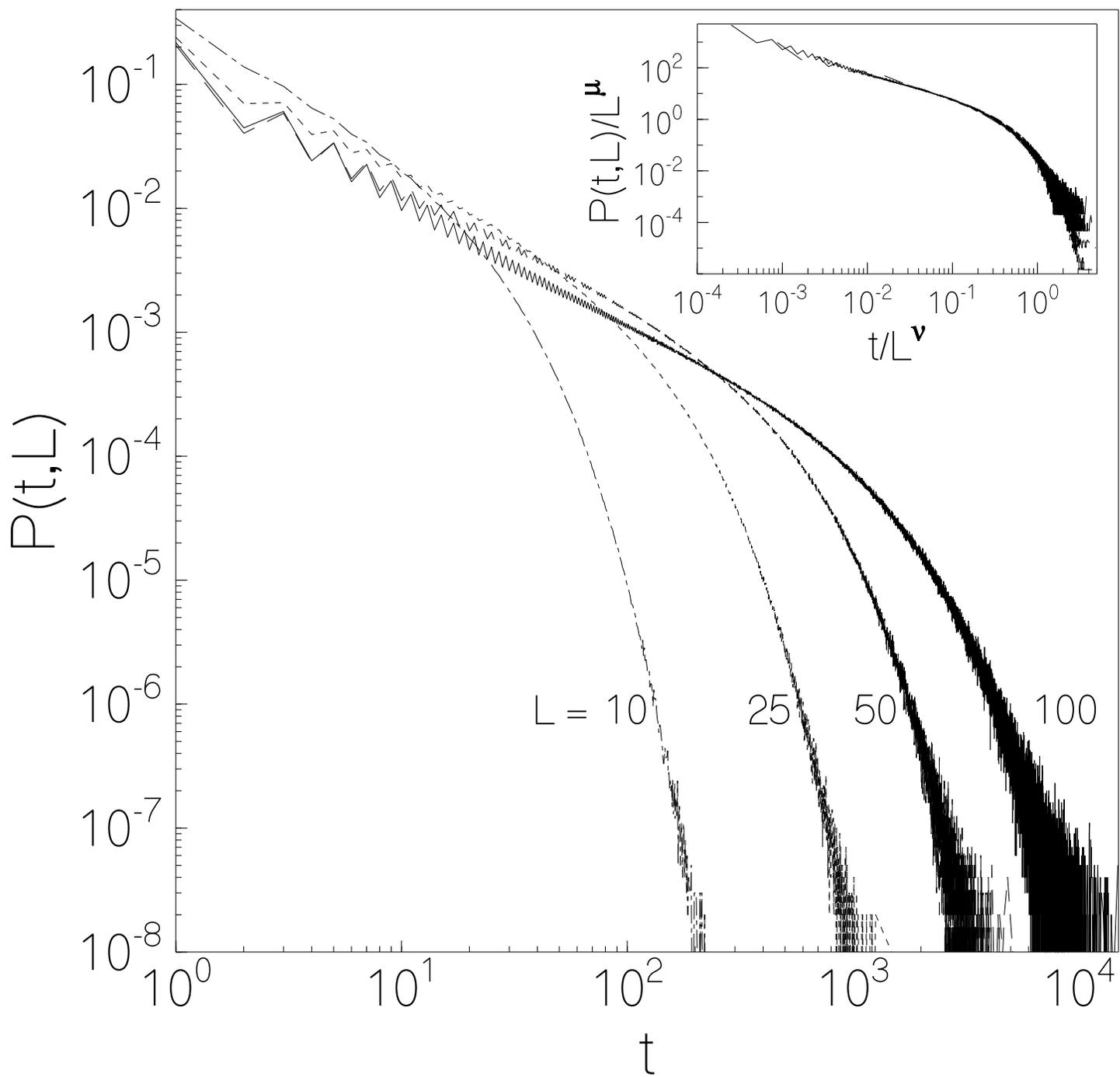
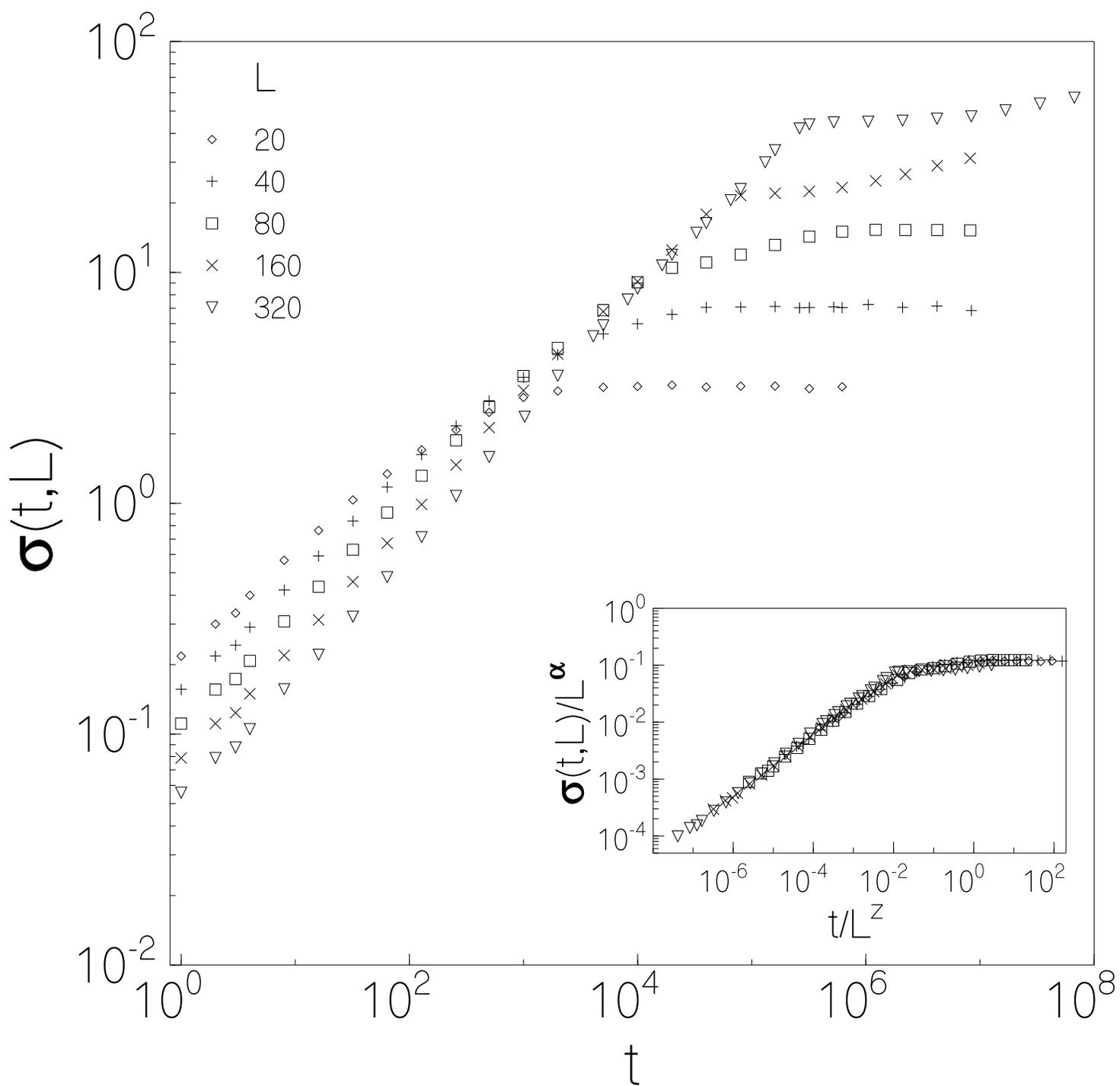
\end{document}